\documentclass[aps,prl,twocolumn]{revtex4}
\usepackage{graphics}
\usepackage{graphicx}
\begin{document}
\title{Hydrogen Embrittlement of Aluminum: the Crucial Role of Vacancies}  
\author{Gang Lu$^{(1)}$ and Efthimios Kaxiras$^{(2)}$}
\affiliation
{
$^{(1)}$Department of Physics, California State University Northridge, 
Northridge, California 91330 \\ 
$^{(2)}$Department of Physics and Division of Engineering and Applied Sciences,\\
 Harvard University, Cambridge, Massachusetts 02138}
\begin{abstract}
We report first-principles calculations which demonstrate that 
vacancies can combine with hydrogen impurities in bulk aluminum and play
a crucial role in the embrittlement of this prototypical ductile solid.    
Our studies of hydrogen-induced vacancy superabundant formation and 
vacancy clusterization in aluminum lead to the conclusion  
that a large number of H atoms (up to twelve) can be 
trapped at a single vacancy, which over-compensates 
the energy cost to form the defect. 
In the
presence of trapped H atoms, three nearest-neighbor single vacancies 
which normally would repel each other, aggregate to form a 
trivacancy on the slip plane of Al, acting  
as embryos for microvoids and cracks
and resulting in ductile rupture along the these planes.
\end{abstract}
\maketitle

Hydrogen degradation of the structural properties of solids,
referred to as embrittlement, 
is a fundamental problem in materials physics.  Despite 
intense studies, the definitive mechanism of H embrittlement in 
metals remains poorly understood.  Four general mechanisms  
have been proposed:
(i) formation of a hydride phase;
(ii) enhanced local plasticity;  (iii) grain boundary weakening and 
(iv) blister and bubble formation \cite{myers}.  The underlying atomic
processes and relative importance of the four mechanisms remain
uncertain, and it is likely that a combination of these processes may contribute to 
embrittlement simultaneously. For these mechanisms to be operational, however, 
a critical  
local concentration of H is required, either to form a hydride phase or 
to initiate cracking at microvoids and grain boundaries. One of the outstanding
problems in the current theories of hydrogen embrittlement is the lack of a 
comprehensive and coherent atomistic mechanism to account for the critical H 
concentrations at crack tips. Moreover, it
is widely observed that H-enhanced dislocation mobility is a prelude to the
embrittlement and that the fracture planes coincide with the slip plane of the
material, which is not the typical situation \cite{myers}; 
how all these phenomena come about still remains a mystery. 
It is generally believed that dislocations are central to 
H embrittlement phenomena, and a large body of work has been dedicated
to elucidate hydrogen-dislocation interaction and its consequences on
embrittlement \cite{myers,birnbaum}. 
Vacancies, being ubiquitously present in solids and having the ability to 
act as impurity traps, could play a central role 
in the embrittlement process, but detailed arguments about this role or  
estimates of its relative importance are totally lacking. 

Recent experiments on H-metal systems offer clues 
on the role that vacancies may play in H embrittlement.  
One set of experiments has 
established that H could induce superabundant vacancy formation in a number of 
metals, such as Pd, Ni, Cr, etc.\cite{fukai1,fukai2}.  
The estimated vacancy 
concentration, C$_V$, in these systems can reach a value as high as 23 at.\% 
\cite{fukai1}.  A conclusion drawn from  
these experiments is that H atoms, originally at interstitial positions in the bulk, 
are trapped at vacancies
in multiple numbers with rather high binding energies. It was speculated that
several (three to six) H atoms can be trapped by a single vacancy, with the 
highest number (six) corresponding to the number of octahedral sites 
around a vacancy in either the 
fcc or the bcc lattice \cite{fukai1}. Actually, we shall show below 
based on first-principles 
theoretical calculations that in Al, the prototypical simple metal and 
ductile solid,
up to {\it twelve} H atoms can be trapped at a single vacancy site. 
The consequence of H trapping is that 
the formation energy of a vacancy defect is lowered by a significant amount, an 
energy that we define as the H trapping energy. Such reduction in the vacancy 
formation energy could result in drastic increase (10$^7$ fold for Fe) of 
equilibrium vacancy concentrations \cite{ohno}. 
The superabundant vacancy formation in turn provides more 
trapping sites for H impurities, 
effectively increasing the apparent H solubility in metals by many orders 
of magnitude. 
For example, it was observed experimentally that about 1000 atomic parts 
per million 
(appm) of H atoms can enter Al accompanied by vacancy formation at
the surface under aggressive H charging conditions, which should be 
contrasted with 
the equilibrium solubility of H in Al of about 10$^{-5}$ appm at
room temperature where the experiments were carried out\cite{birnbaum2}; 
this is a staggering change of {\it eight} orders of magnitude in 
concentration. 
It was futher observed that the H-vacancy defects clustered and 
formed platelets 
lying on the \{111\} planes, which directly lead to 
void formation 
or crack nucleation on the \{111\} cleavage planes \cite{birnbaum2}.
        
In order to elucidate the complex nature of H-vacancy interaction 
and to shed light
on experimental results, we have performed 
first-principles
calculations to examine the energetics and electronic structure 
for the relevant 
H-vacancy complexes in Al. Due to the extremely low solubility of 
H in bulk Al, 
experiments are usually difficult and results are dependent 
on H charging conditions; for such systems, 
first-principles calculations
are particularly useful to complement experimental approaches. 
Our first-principles calculations are based on density functional theory with the 
VASP implementation \cite{vasp} and ultra-soft pseudopotentials \cite{vanderbilt}. 
The local-density approximation (LDA) is used in all of our calculations, 
with checks based on the generalized gradient approximation (GGA) for selected cases.
For Al, we find that
LDA results are consistently closer to experimental values than GGA results, so here we
will rely mainly on LDA numbers to draw physical conclusions. 
We employ a supercell containing 
108 atomic sites in a simple cubic lattice to model bulk Al, 
with a $4\times 4\times 4$ reciprocal space grid in the suprecell 
Brillouin zone and a plane-wave kinetic energy cutoff of 220 eV for the Al-H system. 
With these parameters, we obtain 
the formation energy of a single vacancy (0.66 eV) and the binding energy 
for the nearest-neighbour (NN) divacancy in pure Al (-0.06 eV)  
in excellent agreement with other theoretical \cite{carling} and experimental results
\cite{ehrhart} (Table I). 
We note that the NN divacancy formation energy is negative, implying 
that it is unstable compared to two isolated single vacancies. 
This counter-intuitive 
result is due to charge redistribution in the neighborhood of the vacancy,
which has been interpreted as formation of 
directional covalent/metallic bonds that stabilizes the single vacancy 
configuration against the formation of the divacancy \cite{carling,uesugi}. 
 
\begin{figure}
\includegraphics[height=2.0in]{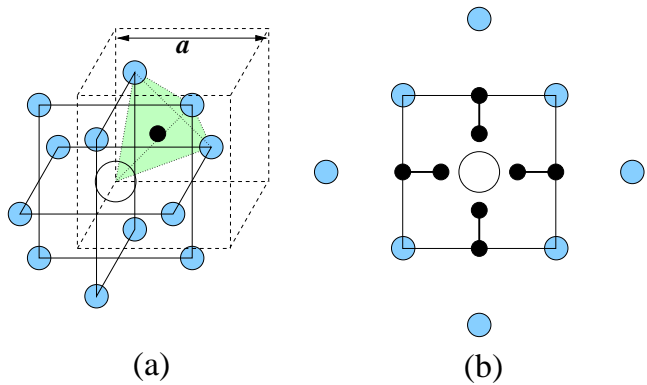}
\caption{
Schematic representation of the environment of a vacncy in Al.
(a) The vacancy as a large open circle and its 12 nearest neighbors as smaller 
grey circles, which lie on highlighted [100] planes.  The cube in dashed lines represents
the conventional cell of the FCC lattice of side $a$.  A shaded tetrahedron with one
corner at the vacancy site is also shown, and the geometric center of which 
corresponds to the lowest-energy site for a H interstitial atom (shown as black circle)
in bulk Al.
(b) The arrangment of four of the six H$_2$ molecules surrounding the vacancy, on a [100] 
plane, with the first- and second-nearest-neighbor Al atoms indicated.
The other two molecules lie directly above and below the plane of the figure, along
an axis perpendicular to this plane passing throught the vacancy site.
In both (a) and (b) the ions are placed at the ideal lattice sites, with the atomic relaxations
not shown explicitly.
}  
\label{fig1}
\end{figure}

Our main objective is to understand the atomistic 
mechanisms of H-vacancy interaction in Al. First we address the 
relative site preference of H in bulk Al.
To this end, we have calculated the total energy of a single H atom 
situated near the vacancy site, or at 
interstitial tetrahedral and octahedral bulk sites which are as far as possible from the 
vacancy within the supercell. 
For H atoms, 
the tetrahedral interstitial site in bulk Al is slightly more favorable than the octahedral interstitial 
site by 0.07 eV. 
We find that the H atom prefers to occupy the vacancy site over the interstitial tetrahedral  
site in bulk by 0.40 eV. 
The corresponding 
experimental value is 0.52 eV \cite{myers}, and theoretical results range from 
0.33 eV to about 1 eV \cite{devita,wolverton}. 
The lowest energy
position for the H atom in the presence of a vacancy 
is {\it not} at the geometric center of the vacancy site, but rather 
at an off-center position close to a tetrahedral site adjacent to the vacancy site (see Fig. 1(a));
the energy difference between the center and off-center positions is 0.66 eV.   
We also find that the H atom is negatively charged, consistent with the view 
that the H impurity can 
be regarded as a screened H$^-$ ion in free-electron-like metals 
\cite{norskov79}. 
Previous studies 
based on the jellium model of Al 
have shown that as the jellium conduction electron density decreases, the 
excess charge buildup 
at the H atom is also reduced and the electrons of the H$^{-}$ ion are less 
localized \cite{norskov77,norskov79}. Therefore, the kinetic energy of the 
H$^{-}$ electrons is lowered at the vacancy site where the conduction electron 
density is lower. At the same time, 
it is energetically favorable for the H$^{-}$ ion to sit off-center of the vacancy,
to minimize the Coulomb interaction energy with the nearby Al ions. 

Having established the stability of a single H atom at a single vacancy in Al, 
the ensuing question
is whether multiple H atoms, in particular, H$_2$ molecules would be stable at 
this defect. 
This is an interesting problem on its own right, but it is also relevant to 
H$_2$ bubble formation that gives rise to H embrittlement. To examine the 
stability of an H$_2$ molecule 
at a vacancy site, we compare the binding energy of the H$_2$ unit at the vacancy 
and in vacuum. The binding energy $E_b$ of the H$_2$ unit at a vacancy 
site is calculated as: 
\begin{equation}
E_b=E_c(V_{\rm Al}+{\rm H}_2)+E_c(V_{\rm Al})-2E_c(V_{\rm Al}{\rm H}),
\end{equation}
where $E_c(V_{\rm Al}+{\rm H}_2)$ is the cohesive energy of a system 
with an H$_2$ unit at the center of the vacancy, 
$E_c(V_{\rm Al})$ is the cohesive energy of a system with a single vacancy 
in the absence of the H$_2$ unit, and $E_c(V_{\rm Al}{\rm H})$ is the cohesive 
energy of a system with a single H atom at the 
vacancy (in the off-center tetrahedral site). Interestingly, we find the 
this binding energy to be
+0.06 eV, indicating a weak repulsion between the two H atoms in the 
H$_2$ unit at the vacancy site. This is to be compared to the
binding energy of an H$_2$ molecule in vacuum, which is $-6.67$ eV. 
The positive binding energy of H$_2$ at the vacancy site 
does not imply that there is no bonding between the two H atoms; it simply 
states that these two H atoms would prefer to be trapped at two single vacancy 
sites {\it individually} rather than in the same vacancy site {\it as a pair}. 
The weakening of the H-H bond at the 
vacancy site is remarkable given the fact that each H atom 
in the H$_2$ unit in this situation
is quite far away (2.6 \AA) from the nearest Al ions. 

We find that the equilibrium interatomic distance between 
the H atoms at the vacancy is 0.83 \AA, 12\% longer than the H$_2$ bond length 
(0.74 \AA) of the molecule in vacuum; this is due to the partial occupation 
of antibonding states between the H atoms. This can be understood as follows: 
each H atom is associated with a doubly occupied 
bound state in the presence of conduction electrons, and hence is negatively charged. 
When the two H atoms approach each other, the two bound states split up into a 
bonding and an antibonding level. In contrast to what happens in vacuum, 
the screening of the conduction electrons reduces the bonding-antibonding energy 
splitting, and the antibonding level may be occupied by conduction electrons 
if the Fermi energy of the metal is high enough \cite{norskov77}. The 
occupation of antibonding states weakens the H$_2$ bond
and increases the bond length. Our results agree qualitatively with the jellium 
model calculations which also found the H$_2$ binding energy to be positive and 
the bond length increased, ranging from 0.81 to 0.86 \AA~ depending on the 
jellium density. In particular, for low jellium electron density 
(corresponding to the center of a vacancy site in Al), the binding energy was 
found to be +0.02 eV \cite{bonev}. 
Similarly, one can calculate the binding energy of multiple H atoms trapped at a 
single vacancy site, which turn out to be positive 
as well. 
Based on these results, we conclude that if the single 
vacancy concentration 
C$_V$ is greater than the H concentration C$_H$, each vacancy in equilibrium 
should contain no more than one H atom. 

\begin{figure}
\includegraphics[width=220pt]{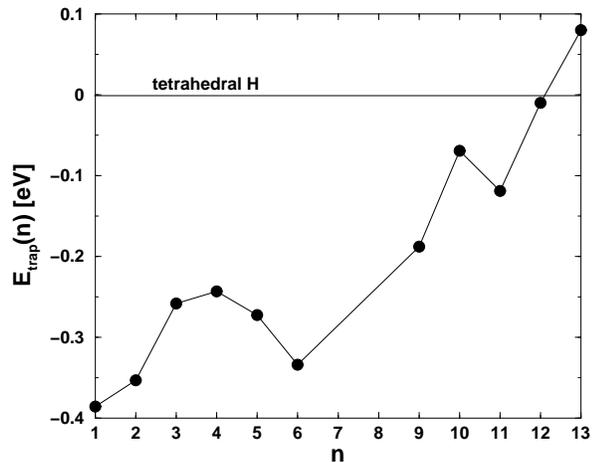}
\caption{Trapping energy per H atom in eV as a function of the number of H atoms
being trapped at a single vacancy site. The zero energy corresponds to the energy 
of a H atom at the tetrahedral interstitial site.}  
\label{fig2}
\end{figure}

On the other hand, if C$_H$ is greater than C$_V$, the question arises 
as to where will the extra H atoms be situated, at  
interstitial or at vacancy sites? 
Experimental measurements for the ratio C$_H$/C$_V$ in Al range from 
0.25 to 4, depending on H charging conditions, with the most probable value 
close to 1 \cite{birnbaum2}.  
To answer the above question, we
have calculated the trapping energy $E_{trap}$ of multiple H atoms at a single vacancy site, 
which is defined as:
\begin{equation}  
E_{trap}(n)=\frac{1}{n}[E_c(V_{\rm Al}+n{\rm H})-E_c(V_{\rm Al})]-
[E_c^0({\rm H})-E_c^0],
\end{equation} 
where $E_c(V_{\rm Al}+n{\rm H})$ is the cohesive energy of a system 
with $n$ H atoms each situated at a single vacancy site,
$E_c^0({\rm H})$ is the cohesive energy of bulk Al with a H atom at the 
tetrahedral interstitial site,
and $E_c^0$ is the cohesive energy of the ideal bulk without H. A negative 
value for the trapping energy 
represents the energy gain when the H atoms are trapped at a single vacancy 
site relative to being dispersed at $n$ different tetrahedral interstitial sites. 
The results for $E_{trap}$ as a function of $n$ 
are summarized in Fig. 2. Consistent with the binding energy calculations, 
it is energetically most favorable for each vacancy to trap a single H atom.
At the same time, it is also energetically favorable for multiple H 
atoms to be trapped at a single vacancy site relative to being dispersed at 
interstitial sites as individual atoms. In fact, up to {\em twelve} H atoms can be trapped at a 
single vacancy in Al, twice the highest number of H atoms (six) that can 
be trapped in Fe \cite{ohno}. 
The atomic arrangement of the 12 H atoms trapped at a single vacancy is indicated 
in Fig. 1(b). There are two H$_2$ units in each 
$\langle$100$\rangle$ direction surrounding the vacancy, and the bond length is 
1 \AA~ for all six units.
The inter-molecule distance in each direction is 3 \AA, and the NN
distance between H and Al in each direction is 2 \AA (the lattice constant of 
Al is 3.99 \AA). The ordered arrangement of the H atoms is necessary to 
minimize the electrostatic energy. The greater H-trapping capacity of Al 
compared to Fe, can be attributed to its larger lattice constant and 
more delocalized nature of electrons. It is observed that the volume change 
of the supercell owing to the H additions is negligible.  

The fact that the H$_2$ units at a single vacancy site attract the conduction 
electrons from the edge of the vacancy, raises the interesting possibility 
that the covalent/metallic bonds 
between the first shell of NN Al ions around the vacancy site 
may be disrupted enough to permit a coalescence of 
multiple vacancies. To check this possibility, we carried out calculations 
for a number of relevant configurations. Specifically, we have
examined: (i) two vacancies, each with one H atom, forming a NN divacancy 
with two H atoms trapped; (ii) $n$ vacancies, each with two H atoms, forming 
a complex of NN multi-vacancies with $2n$ H atoms trapped, for $n=2$ and 3. 
To summarize the results, we use the notation of chemical reactions: 
\begin{eqnarray*}
2 \; V_{\rm Al}{\rm H} & \rightarrow &  
(V_{\rm Al})_2{\rm H}_2 - 0.21\; 
{\rm eV} 
\; \; \; 
({\rm i}) 
\\
n \; V_{\rm Al}{\rm H}_2 & \rightarrow &  
(V_{\rm Al})_n{\rm H}_{2n} + n \; 0.29\; 
{\rm eV} 
\; \; \; 
({\rm ii}) 
\end{eqnarray*}
where the last number in each equation represents the reaction enthalpy 
$\Delta H$.  
A positive value of $\Delta H$ means the reaction is exothermic, that is, the 
process from left to right is energetically favorable. 
$\Delta H$ is defined as follows for 
reaction (i):
\begin{equation}
\Delta H=2E_c(V_{\rm Al}{\rm H})-E_c[(V_{\rm Al})_2{\rm H}_2]-E_c^0,
\end{equation}
where $E_c[(V_{\rm Al})_2{\rm H}_2]$ is 
the cohesive energy of a system with two H atoms trapped at a 
divacancy, with analogous definitions for reaction (ii). 
Consistent
with our earlier discussion, we find reaction (i) to be unfavorable (endothermic) because 
the effect of a {\it single} H atom on the covalent/metallic bonding of the 
NN Al atoms around the vacancy site is small and localized. 
On the other hand, reaction (ii) is 
favorable for $n=2$ and 3, because the H$_2$ units can attract more conduction electrons from the 
nearby Al atoms, weakening the bonding among the NN Al atoms, which in turn drives
the formation of multi-vacancies. 
The large energy gain in forming the trivacancy ($n=3$) is of particular interest. 
First, it is consistent with the experimental observation that the single 
vacancy defects occupied by H atoms can 
coalescence to form platelets on \{111\} planes of Al. Although our 
calculations primarily concern the formation of the trivacancy, 
it is likely that even larger vacancy clusters can also be formed based on 
the same mechanism.  In support of this claim, we mention that 
the increase in positive enthalpy associated with
reaction (ii) is linear in the number of vacancies for $n=2$ and 3. 
Second, these vacancy clusters
can serve as embryos of cracks and microvoids with local H concentrations much 
higher than the average bulk value.  

Next we discuss the implications of our results on hydrogen embrittlement 
phenomena. It was generally believed that H-induced embrittlement in 
metals takes the form of plastic rupture rather 
than brittle fracture, consistent with the notion of H-enhanced local 
plasticity (HELP). It was widely observed that the fracture surface is 
along the active slip planes where shear 
localization occurs. For fcc metals, the slip planes are the \{111\} planes. 
In many cases, 
microvoids open up along these active slip planes in front of the crack tip; 
these microvoids can open and close in response to the local stress. 
Plastic rupture occurs when these microcracks are joined to the crack tip, 
upon reaching the critical stress.
Our results clearly suggest that H-enriched microvoids may be created along the 
slip planes by the coalescence of vacancies with trapped H. These microvoids 
can be formed 
{\it only} in the presence of H, which produces an {\it additional} source of 
microcracks necessary for the H embrittlement. Moreover, the H-induced 
vacancy formation also facilitates 
dislocation climb, leaving behind vacancy rows in the highly deformed regions, 
which may contribute to the formation of microcracks as well. 
Our studies, taken together with the observed vacancy-enhanced 
dislocation glide \cite{lauzier,benoit}, suggest that vacancies are also 
responsible 
for the HELP phenomena that are a prelude to H embrittlement \cite{myers}. 
The fact that there is a strong binding between H and dislocation cores, 
and H can enhance dislocation motion along the slip planes \cite{lu}, 
provide a means of rapid transport of H atoms to the crack front.  
On the other hand, the apparent 
lattice mobility of H atoms is also enhanced since multiple H atoms may be 
trapped at a single vacancy. All these vacancy-based mechanisms contribute 
to the H embrittlement as they increase the
rate of crack growth. Finally, the significant H trapping at vacancies 
provides a scenario by which drastic increase of local H concentration may 
occur without improbable accumulation of H at bulk interstitial sites\cite{ohno}. 
This new feature resolves the long-standing problem 
of how a sufficiently high H concentration can be realized to 
successfully induce H embrittlement in 
materials such as Al, where the equilibrium H concentration in bulk is 
extremely low. 

We acknowledge the support from Grant No. F49620-99-1-0272
through the U.S. Air Force Office for 
Scientific Research.

\begin{table}[t]
\caption{The vacancy formation energy, $\Delta H^F_V$; the binding energy for
the divacancy, $\Delta H^b_{2V}$=2$\Delta H^F_V$ - $\Delta H^F_{2V}$, where
the last term is the formation energy of the divacancy. The total energy with a
H atom occupying the octahedral interstitial site is set to zero, relative to which
the total energy of a H atom occupying the tetrahedral interstitial site, E$_T$, and the 
total energy of a H atom trapped at a single vacancy, E$_V$ are defined. The
last two columns are LDA and GGA results from other theoretical calculations.
All energies are given in eV. The experimental values marked by an asterisk
have been called into question due to incorrect interpretations on the
experimental part,
see ref. \cite{carling} for details.}
\begin{ruledtabular}
\begin{tabular*}{\columnwidth}{@{\extracolsep{\fill}}cccccc}
                    & LDA & GGA  &  Exp.\cite{ehrhart} & LDA & GGA \\ \hline
 $\Delta H^F_V$    & 0.66  & 0.54 & 0.67$\pm$0.03 & 0.70 \cite{carling} & 
 0.54  \cite{carling} \\
 $\Delta H^b_{2V}$    & -0.06  & -0.07 & 0.2$^{\ast}$, 0.3$^{\ast}$ 
& -0.07 \cite{carling}& -0.08  \cite{carling}\\
 E$_T$  & -0.07 &  & & -0.05 \cite{wolverton} & -0.13  \cite{wolverton} \\
 E$_V$  & -0.47 &  & & &-0.46 \cite{wolverton} \\
\end{tabular*}     
\end{ruledtabular}        
\label{table1}
\end{table}


\begin{thebibliography}{99}
\bibitem{footnote}
Using the energy curve between equilibrium and metastable 
  positions of H in bulk Al, we estimated the tunneling probability
  between equivalent equilibrium positions to be 
 $3\times 10^{-8}$, and thus quantum effects are not expected 
  to be significant.  Thermal effects (the configurational and 
  vibrational entropy) are well beyond the reach of first-principles
  calculations for this system.
\end{thebibliography}
\end{document}